\documentclass[conference]{IEEEtran}

\def\BibTeX{{\rm B\kern-.05em{\sc i\kern-.025em b}\kern-.08emT\kern-.1667em\lower.7ex\hbox{E}\kern-.125emX}}    

\usepackage{blindtext}
\usepackage{graphicx}
\usepackage{comment}
\usepackage{amsmath,amssymb}
\usepackage{url}
\usepackage{fancyhdr}
\usepackage{cite}
\usepackage[normal]{subfigure}
\usepackage{color}
\usepackage{acronym}
\usepackage{comment}
\usepackage{float}
\usepackage{algorithm}
\usepackage[utf8]{inputenc} 
\usepackage[T1]{fontenc}
\usepackage{ifthen}
\usepackage{cite}
\usepackage{multirow}
\usepackage[table]{xcolor} 

\acrodef{CIR}{channel impulse response}
\acrodefplural{CIR}[CIRs]{channel impulse responses}
\acrodef{CSI}{channel state information}
\acrodef{ML}{maximum likelihood}
\acrodef{LLR}{logarithm of the likelihood ratio}
\acrodef{MIMO}{multiple-input multiple-output}
\acrodef{OFDM}{orthogonal frequency division multiplexing}
\acrodef{c.d.f.}{cumulative density function}
\acrodef{GLRT}{generalized likelihood ratio test}
\acrodef{SNR}{signal-to-noise ratio}
\acrodef{AWGN}{additive white Gaussian noise}
\acrodef{TPR}{true positive rate}
\acrodef{TNR}{true negative rate}
\acrodef{NN}{Nearest Neighbor}
\acrodef{OCC}{One-class classification}
\acrodef{PLA}{physical layer authentication}
\acrodef{OCNN}{one-class nearest neighbor}

\usepackage[margins]{trackchanges}
\addeditor{LS}

\begin{document}
%\title{Statistical and Machine Learning-based Decision Techniques for Physical Layer Authentication over Time-Varying Fading Channels}
\title{Statistical and Machine Learning-based Decision Techniques for Physical Layer Authentication}

\author{%
  \IEEEauthorblockN{Linda Senigagliesi, Marco Baldi and Ennio Gambi}
  \IEEEauthorblockA{Dipartimento di Ingegneria dell'Informazione\\
                    Universit\`a Politecnica delle Marche\\
                    60131 Ancona, Italy\\
                    Email:  \{l.senigagliesi, m.baldi, e.gambi\}@univpm.it,        \\ }
}

\maketitle

\begin{abstract}
In this paper we assess the security performance of key-less physical layer authentication schemes in the case of time-varying fading channels, considering both partial and no channel state information (CSI) on the receiver's side.
We first present a generalization of a well-known protocol previously proposed for flat fading channels and we study different statistical decision methods and the corresponding optimal attack strategies in order to improve the authentication performance in the considered scenario. 
We then consider the application of machine learning techniques in the same setting, exploiting different one-class nearest neighbor (OCNN) classification algorithms.
We observe that, under the same probability of false alarm, one-class classification (OCC) algorithms achieve the lowest probability of missed detection when a low spatial correlation exists between the main channel and the adversary one, while statistical methods are advantageous when the spatial correlation between the two channels is higher.

%In this paper we generalize a physical layer authentication protocol previously proposed for flat fading wireless channels, extending its application to the case of time-varying fading channels.
%We assess the loss in performance coming from channel fading as a function of the fading severity.
%We then consider three different decision methods for signal authentication in order to improve the protocol performance in the considered setting.
%We also derive the corresponding optimal attack strategies and assess the security of the protocol in the presence of optimal and sub-optimal attackers.
\end{abstract}

%
% The code below should be generated by the tool at
% http://dl.acm.org/ccs.cfm
% Please copy and paste the code instead of the example below. 

% \begin{CCSXML}
% <ccs2012>
%  <concept>
%   <concept_id>10010520.10010553.10010562</concept_id>
%   <concept_desc>Computer systems organization~Embedded systems</concept_desc>
%   <concept_significance>500</concept_significance>
%  </concept>
%  <concept>
%   <concept_id>10010520.10010575.10010755</concept_id>
%   <concept_desc>Computer systems organization~Redundancy</concept_desc>
%   <concept_significance>300</concept_significance>
%  </concept>
%  <concept>
%   <concept_id>10010520.10010553.10010554</concept_id>
%   <concept_desc>Computer systems organization~Robotics</concept_desc>
%   <concept_significance>100</concept_significance>
%  </concept>
%  <concept>
%   <concept_id>10003033.10003083.10003095</concept_id>
%   <concept_desc>Networks~Network reliability</concept_desc>
%   <concept_significance>100</concept_significance>
%  </concept>
% </ccs2012>  
% \end{CCSXML}

% \ccsdesc[500]{Computer systems organization~Embedded systems}
% \ccsdesc[300]{Computer systems organization~Redundancy}
% \ccsdesc{Computer systems organization~Robotics}
% \ccsdesc[100]{Networks~Network reliability}

% We no longer use \terms command
%\terms{Theory}


\section{Introduction}

The physical transmission layer has gained attention from a security point of view in the last years, thanks to its ability to guarantee secrecy without any need of pre-shared keys \cite{bloch_barros_2011}.
In addition, contrary to classic cryptographic approaches based on computational security, physical layer security metrics do not require to make any assumption on the computing power of attackers.
This, jointly with the low algorithmic complexity of physical layer security techniques, has made this area of research of great relevance for some scenarios like that of resource-constrained wireless devices \cite{Mukherjee2015}.
In fact, the diffusion of the Internet of Things (IoT) and a growing number of devices deployed over networks have forced security and authentication schemes to meet new requirements, such as low latency and delay, limited storage and energy constraints. Traditional cryptographic methods, however, are characterized by a complexity that barely fits these requirements. Physical layer techniques on the other hand help in reducing the burden of heavy computations and guarantee a level of security which does not depend on the attacker's capabilities. 
%\mb{Qui qualche riferimento piu recente (ma comunque autorevole) lo metterei, magari relativo a PHYSEC in IoT. Guardiamo i lavori piu recenti degli autori famosi che conosciamo.}

Authentication aims at recognizing messages coming from a legitimate transmitter, while detecting those originated by a malicious attacker. At the physical layer, authentication is performed exploiting the unique characteristics of the communication channel to distinguish the source of the message~\cite{Caparra2017}.
%Although many works on schemes relying on the sharing of a secret key exist \cite{Xie2018, Choi2019}, we consider a key-less approach, as done in \cite{Xiao2008b,Lai2009}.

In this paper, we assess the security performance achievable by \ac{PLA} protocols over wireless parallel channels affected by time-varying fading. Effects of time-varying fading have been studied in literature considering schemes based both on the sharing of a secret key \cite{Xie2018} and key-less approaches \cite{Xiao2008, Tugnait2014}.
In order to improve the performance degradation due to channel time variability, we propose different decision methods, starting from the work done in \cite{Baracca2012} in a flat fading model. We first consider statistical criteria which follow a hypothesis testing approach \cite{Maurer2000b}, and their corresponding optimal attacks. We then exploit machine learning resources, whose application to \ac{PLA} has started to attract interest in literature in the last years under different scenarios \cite{Weinand2017, Wang2017}. Only recently, authors in \cite{Fang2019} proposed the use of kernel machine-based methods to improve the authentication performance in time-varying environments.
With respect to \cite{Fang2019}, we consider a more realistic training phase corresponding only to the initial state of the channel and the existence of some correlation between the main channel and the adversary's one. We also study \ac{OCNN} algorithms, and show that in the considered setting they achieve good performance, especially when the attacker's channel has a low correlation with the main one. 

The paper is organized as follows. In Section \ref{sec:sys} the system model is described. In Section \ref{sec:dec_methods} we present three different decision methods that can be adopted by Bob and the corresponding optimal attack strategies. Section \ref{sec:class} introduces one-class classification techniques. Numerical results about the security performance of the system are provided in Section \ref{sec:results}. Finally, Section \ref{sec:conclusion} concludes the paper.

%Notation: In the rest of the paper, bold letters denote vectors, $\mathcal{CN} (\mathbf{a}, \mathbf{R})$ represents the distribution of circularly symmetric complex Gaussian random vectors with mean vector $\mathbf{a}$ and covariance matrix $\mathbf{R}$.

\section{System model}
\label{sec:sys}

\begin{figure}[ht]
\centering
\includegraphics[width=0.35\textwidth]{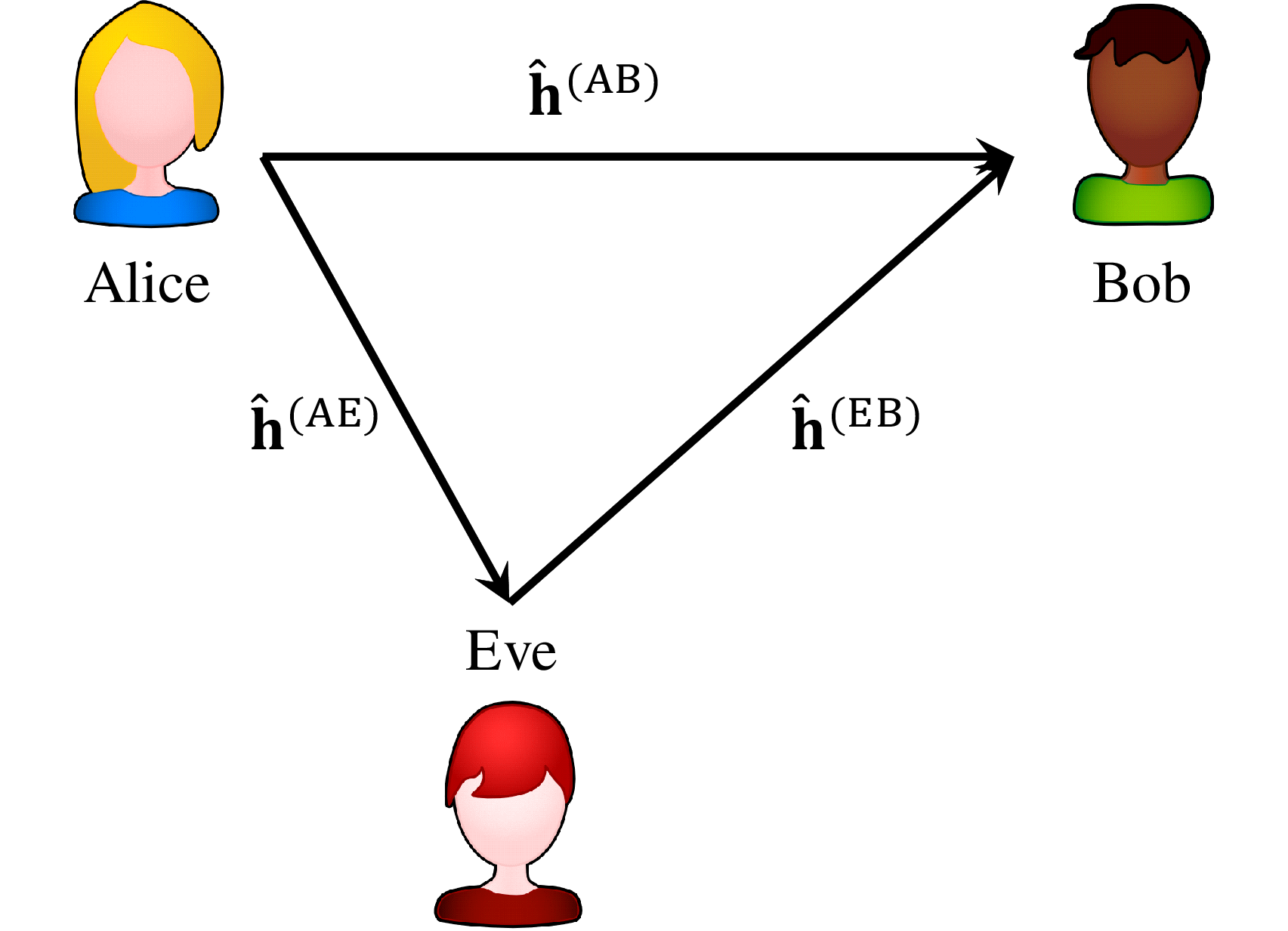}
\caption{Physical layer authentication channel model.\label{fig:Fig1}}
\end{figure}

We analyze the channel model depicted in Fig. \ref{fig:Fig1} where a peer, Alice, has to be authenticated by an authenticator, Bob, at the presence of a malicious attacker, Eve, who aims at impersonating Alice by forging her messages. Through \ac{PLA}, Bob should be able to distinguish the messages coming from Alice as legitimate and to refuse those coming from Eve.
%\mb{Come fatto qui sopra, per gli utenti userei la nomenclatura piu ortodossa dei protocolli di autenticazione, ovvero peer (Alice), autenticator (Bob) e attacker (Eve). Andrebbe pero fatto coerentemente dappertutto.}

A set of $N$ parallel channels is considered, modeling for example a multi-carrier or \ac{OFDM} transmission.
The \acp{CIR} of each channel are collected into a vector $\mathbf{h}$ of complex numbers, whose entries are zero-mean correlated circularly symmetric complex Gaussian variables. The channel between Alice and Bob is denoted as an $N$-size vector
\begin{equation}
\mathbf{h}^{(\textsc{AB})} \sim \mathcal{CN} (\mathbf{0}_{N\times 1}, \mathbf{R}^{(\textsc{AB})}).
\label{eq:hvec}
\end{equation}
Transmissions are subject to slow or fast fading, which deeply affects the quality of authentication, and to \ac{AWGN}. Depending on how fast the channel varies over time, for Bob it is more difficult to establish whether any received packet comes from Alice or from Eve.

The authentication procedure we consider is based only on channel estimates, and does not require additional cryptographic mechanisms. In general, an authentication protocol comprises two phases, that are summarized in the following.

%\paragraph{Phase I} 
\textit{Phase I:}
In the setup phase, Bob receives a known packet that surely comes from Alice. This can be guaranteed either by exploiting higher layer cryptographic schemes or through physical measures (e.g., by manually executing the setup phase).
By exploiting the setup packet (with known content), Bob obtains an estimate of the channel between himself and Alice at time $t_0$, which can be written as
\begin{equation}
    \hat{\mathbf{h}}^{(\textsc{AB})} = \mathbf{h}^{(\textsc{AB})} + \mathbf{w}^{(\textsc{I})},
    \label{eq:2}
\end{equation}
where $\mathbf{w}^{(\textsc{I})} \sim \mathcal{CN} (\mathbf{0}_{N\times 1}, \sigma_I^2\mathbf{I}_N)$ is a noise vector.

%\paragraph{Phase II}
\textit{Phase II:}
During normal operation, Bob receives further packets from Alice and exploits them to estimate the channel between him and Alice again.
By comparing any of these estimates with the reference one obtained during the setup phase, Bob tries to understand whether each packet comes from Alice or not.
In order to evaluate security in the worst case scenario, we suppose that in this phase Eve can potentially modify the channel estimation obtained by Bob into any vector $\mathbf{g}$, as already done in \cite{Baracca2012}. 
%As done in \cite{Baracca2012}, we suppose that in this phase Eve can potentially modify the channel estimation obtained by Bob into any vector $\mathbf{g}$. 

Bob resorts to an hypothesis test \cite{Kay1993} to decide whether the transmission was performed by Alice or not. Denoting by $\hat{\mathbf{h}}$ the channel estimated by Bob, the two hypotheses are
\begin{itemize}
\item $\mathcal{H}_0$: the message is coming from Alice.
This hypothesis can be verified on the basis of the correlation between estimates made in different time instants, which depends on the severity of fading affecting the channel. We represent it as a real number between 0 and 1, where 1 corresponds to the case of a flat fading channel.
We work under the hypothesis of slowly time-varying fading channels, meaning that the fading coefficient is assumed constant during transmission of each packet \cite{Wiesel2006}.
The hypothesis $\mathcal{H}_0$ at time $t$ can therefore be written as
\begin{equation}
    \hat{\mathbf{h}}(t) = \boldsymbol{\alpha}(t)\mathbf{h}^{(\textsc{AB})} + \sqrt{1- \boldsymbol{\alpha}^2(t)}\mathbf{w}_F + \mathbf{w}^{(\textsc{II})}(t) ,
    \label{eq:H0_f}
\end{equation}
where $\boldsymbol{\alpha} = \left[\alpha_1, \cdots, \alpha_n \right]$ represents the vector of time correlations on each channel. %, which take into account the effect of fading during time in each channel; the coefficients $\alpha_i$ are real numbers that belongs to the interval $[0,1]$. The case of $\alpha_i=1$ corresponds to the case of flat fading on the $i$-th channel.
$\mathbf{w}_F$ represents a random variable generated according to a Rayleigh statistics with unitary variance. This is different from \cite[eq. (59)]{Tomasin2018}, where Jakes fading is considered.

\item $\mathcal{H}_1$: the message is coming from Eve, and
\begin{equation}
    \mathbf{\hat{h}}(t) = \mathbf{g} + \mathbf{w}^{(\textsc{II})}(t).
    \label{eq:H1_f}
\end{equation}
\end{itemize}

From Eve's point of view, we suppose that she knows exactly which decision strategy is applied by Bob, and performs a tailored attack. Moreover, we suppose that the attacker has a partial \ac{CSI}, so she knows the statistics of the channels between Alice and Bob and between Alice and herself, but not the exact channel realizations.

We assume that Eve can observe transmissions from Alice to Bob and vice versa, thus estimating $\mathbf{h}^{(\textsc{AE})}$ and $\mathbf{h}^{(\textsc{EB})}$. 
%We assume that Eve can observe transmissions from Alice to Bob and estimate $\mathbf{h}^{(\textsc{AE})}$, but not the channel between her and Bob. Bob in fact receives messages from Alice but is not expected to transmit enough messages to allow the attacker to extract useful information on $\mathbf{h}^{(\textsc{EB})}$. 
We denote the channel estimates performed by Eve at instant $t_0$ as
\begin{align}
   &\hat{\mathbf{h}}^{(\textsc{AE})} = \rho_{\textsc{AE}}\mathbf{h}^{(\textsc{AB})} + \sqrt{1-\rho_{\textsc{AE}}^2}\mathbf{r} + \mathbf{w}^{(\textsc{AE})}, \\
    &\hat{\mathbf{h}}^{(\textsc{EB})} = \rho_{\textsc{EB}}\mathbf{h}^{(\textsc{AB})} + \sqrt{1-\rho_{\textsc{EB}}^2}\mathbf{r} + \mathbf{w}^{(\textsc{EB})}, 
\label{eq:Eve_est}
\end{align}
where $\mathbf{w}^{(\textsc{AE})} \sim \mathcal{CN} (\mathbf{0}_{N\times 1}, \sigma_{\textsc{AE}}^2\mathbf{\textsc{I}}_N)$ and $\mathbf{w}^{(\textsc{EB})} \sim \mathcal{CN} (\mathbf{0}_{N\times 1}, \sigma_{\textsc{EB}}^2\mathbf{\textsc{I}}_N)$ represent the noise vectors and $\mathbf{r}$ is a random component with complex entries. $\rho_{\textsc{AE}}$ and $\rho_{\textsc{EB}}$ denote the spatial correlation between the attacker channels (Alice-Eve and Eve-Bob) and the main channel, respectively. 
%\textcolor{blue}{Since Bob relies only on channel estimates collected during Phase I, it is reasonable to assume for Eve to base her attack only on estimates overheard in the same phase, and not on the successive ones, which are affected by time-varying fading.}
%We suppose that the channels between Alice and Eve  and Eve and Bob have a spatial correlation with the channel between Alice and Bob denoted as $\rho_{\textsc{AE}}$ and $\rho_{\textsc{EB}}$ respectively.

We suppose that Eve's attack is based only on the estimates overheard during Phase I, since Bob relies only on this for the subsequent authentication phase.
This is opposed to a differential authentication approach \cite{Tomasin2018}, in which Bob progressively updates its reference estimate.
The latter, however, is suitable in the case of correlated fading over time, which is not the case we consider.
As a consequence, the time coefficient vector $\boldsymbol{\alpha}(t)$ does not affect her forged vector $\mathbf{g}$.

\section{Statistical decision techniques}
\label{sec:dec_methods}
In this section we discuss two different statistical criteria for Bob to decide between the two hypotheses $\mathcal{H}_0$ and $\mathcal{H}_1$.
We evaluate the reliability of these tests by measuring the errors that follow from them, i.e. the probability of false alarm (FA) and the probability of missed detection (MD).
By \textit{false alarm} we mean the event that occurs when Bob rejects a message coming from Alice, while there is a \textit{missed detection} when he accepts a message forged by Eve.
If we take a measure to reduce the probability of a FA, normally this increases the probability of a MD.
Therefore, a trade-off between these two effects has to be found.

It is important to observe that while time-varying fading negatively affects the correct authentication of legitimate signals, it plays a positive role from the attacker's point of view. As will be shown in the following, it directly influences the probability of FA, forcing Bob to accept a larger range of inputs, and thus increasing the chances for Eve that one of her forged messages is accepted as authentic.

\subsection{Logarithm of likelihood ratio test}

Let us first examine the \ac{GLRT} \cite{Kay1993}, as done in \cite{Baracca2012}, where flat fading channels were considered. We take into account the more general case in which channel variations occur during the authentication.
%Please the reader refers to \cite{Baracca2012} for a detailed description of 
%In this case, Eve chooses $\mathbf{g}$ the corresponding maximum likelihood (ML) estimate, and the resulting vector estimated by Bob is $\mathbf{\hat{h}}$ from \eqref{eq:H1_f}.

The \ac{GLRT} consists in comparing the \ac{LLR} with a threshold $\theta > 0$, i.e.
\begin{equation}
 \begin{cases}  \Psi \leq \theta : & \mbox{decide for }\mathcal{H}_0, \\ 
\Psi > \theta : & \mbox{decide for }\mathcal{H}_1.
\end{cases}
\label{eq:threshold}
\end{equation}
According to \cite{Baracca2012}, the LLR can be written as
\begin{equation}
    \Psi \varpropto 2 \sum_{n=1}^{N} {\frac{1}{\sigma_n^2}\left|\hat{h}_n - \hat{h}_n^{(\textsc{AB})} \right|^2},
    \label{eq:psi}
\end{equation}
where $\sigma_n^2$ represents the per-dimension variance, evaluated as $\sigma_n^2 = \sigma_{\textsc{I}}^2 + \sigma_{\textsc{II}}^2 + {1-\alpha_n^2(t)}$.

By substituting \eqref{eq:H0_f} in \eqref{eq:psi}, we obtain that under the hypothesis $\mathcal{H}_0$, $\Psi$ is a non-central chi-square random variable as in \cite{Baracca2012}, with non centrality parameter
%\begin{equation}
%    \mu = \frac{1}{\sigma^2}\left\|(\alpha-1)\mathbf{h}^{(\textsc{AB})}\right\|^2.
%    \label{eq:mu}
%\end{equation}
\begin{equation}
    \mu = \sum_{n=1}^N \frac{1}{\sigma_n^2}\left|(\alpha_n-1)\mathbf{h}^{(\textsc{AB})}\right|^2.
    \label{eq:mu}
\end{equation}
We note that $\mu$ is strictly dependent on $\boldsymbol{\alpha}$, and becomes zero in the limit case of $\alpha_n = 1$ on each channel (absence of fading), which boils down to the case shown in \cite{Baracca2012}.

We can evaluate the probability of false alarm $P_{\textsc{FA}}$, i.e. the probability that Bob refuses a message coming from Alice, as
\begin{equation}
    P_{\textsc{FA}} = P[\Psi>\theta|\mathcal{H}_0] = 1 - F_{\chi^2,\mu}(\theta),
    \label{eq:P_FA}
\end{equation}
where $F_{\chi^2,\mu}(\cdot)$ denotes the \ac{c.d.f.} of a chi-square random variable with $2N$ degrees of
freedom and noncentrality parameter $\mu$.

By substituting the hypothesis $\mathcal{H}_1$ in the \ac{LLR} \eqref{eq:psi}, we note that $\Psi$ is again a non-central chi-square random variable, but the noncentrality parameter in this case is
%\begin{equation}
%    \beta = \frac{1}{\sigma^2}\left\|\mathbf{g}-\mathbf{h}^{(\textsc{AB})}\right\|^2.
%    \label{eq:beta}
%\end{equation}
\begin{equation}
    \beta = \sum_{n=1}^N \frac{1}{\sigma_n^2}\left|\mathbf{g}-\mathbf{h}^{(\textsc{AB})}\right|^2.
    \label{eq:beta}
\end{equation}
We can calculate the probability of missed detection as
\begin{equation}
    P_{\textsc{MD}} = P[\Psi \leq \theta|\mathcal{H}_1] =  F_{\chi^2,\beta}(\theta).
    \label{eq:P_MD}
\end{equation}
By imposing a target $P_{\textsc{FA}}$, the threshold is set as
\begin{equation}
    \theta = F_{\chi^2,\mu}^{-1}(1-P_{\textsc{FA}}).
    \label{eq:thr}
\end{equation}

For the first decision method we resort to the attack strategy developed in \cite{Baracca2012}. %We say that Bob accepts a message if its corresponding channel estimate $\hat{\mathbf{h}}$ lies inside the sphere $\mathcal{S}$ (in the $N$-dimensional complex space $\mathbb{C}_N$) centered around $\hat{\mathbf{h}}^{(\textsc{AB})}$ and having radius $r = \sqrt{\frac{\theta}{2}\sigma_n^2}$. Since $\sigma_n^2 = \sigma_{\textsc{I}}^2 + \sigma_{\textsc{II}}^2 + \sqrt{1-\alpha_n^2(t)}$, the radius dimension increases with the severity of fading, leading Bob to accept more messages and increasing the probability of a successful attack. 
The maximum probability of attack success is achieved with the \ac{ML} estimate of $\hat{\mathbf{h}}^{(\textsc{AB})}$ based on the observations $\hat{\mathbf{h}}^{(\textsc{AE})}$ and $\hat{\mathbf{h}}^{(\textsc{EB})}$ available to Eve.
From \cite[eq. (45)]{Baracca2012} we obtain
%\begin{equation}
%    g_n = \rho_{\textsc{AE}}\hat{h}_n^{(AE)}.
%\label{eq:att1}
%\end{equation}

\begin{equation}
    g_n = \hat{h}_n^{(\textsc{EB})}C_n + \hat{h}_n^{(\textsc{AE})}D_n ,
\label{eq:att1}
\end{equation}
writing $C_n$ and $D_n$ as
\begin{subequations}
\begin{equation}
C_n = \frac{\rho_{\textsc{EB}}\omega_n^{(\textsc{EB})} - \rho_{\textsc{AB}}\rho_{\textsc{AE}}}{\omega_n^{(\textsc{AE})}\omega_n^{(\textsc{EB})}-\rho_{\textsc{AB}}^2},
    \label{eq:Cn}
\end{equation}
\begin{equation}
D_n = \frac{\rho_{\textsc{AE}}\omega_n^{(\textsc{AE})} - \rho_{\textsc{AB}}\rho_{\textsc{EB}}}{\omega_n^{(\textsc{AE})}\omega_n^{(\textsc{EB})}-\rho_{\textsc{AB}}^2},
    \label{eq:Dn}
\end{equation}
\end{subequations}
where $\omega_n^{(\textsc{AE})} = 1 + \frac{\sigma_{\textsc{AE}}^2}{\lambda_n}$, $\omega_n^{(\textsc{EB})} = 1 + \frac{\sigma_{\textsc{EB}}^2}{\lambda_n}$ and $\lambda_n$ represents the power delay.
In the following we denote the choice \eqref{eq:att1} as \textit{LLR attack} for brevity.

\subsection{Combined test}
%The \ac{LLR} test alone however is not sufficient to guarantee a correct authentication (and a small probability of MD) for values of $\mathbf{\alpha}$ that are not next to $1$. 

When the authentication metric is the sole \ac{LLR}, Eve's forged signal can be optimized according to \eqref{eq:att1}.
If we change the test used by Bob, we naturally improve the authentication  performance under the assumption that Eve's is unaware of Bob's test.
If instead Eve has perfect knowledge of Bob's acceptance strategy, then the same conclusion could no longer hold true.
In order to elaborate on this aspect, let us consider an alternative decision strategy for Bob, based on a double verification.
Introducing a second decision metric besides the \ac{LLR} provides a more selective criterion for Bob to authenticate Alice.
As an example of second metric to be added to the \ac{LLR}, let us consider the modulus of the channel estimates.
%\textcolor{blue}{The latter metric is related to the former one, but provides a more restrictive decision criterion due to the triangle inequality
%$\left\Vert\hat{\mathbf{h}} \right\Vert - \left\Vert\hat{\mathbf{h}}^{(\textsc{AB})} \right\Vert \le \left\Vert\hat{\mathbf{h}} - \hat{\mathbf{h}}^{(\textsc{AB})} \right\Vert$. (IO QUESTA PARTE LA TOGLIEREI)}
The additional test based on the modulus is performed by comparing the modulus of the reference estimate $\hat{\mathbf{h}}^{(\textsc{AB})}$, which represents the only information available to Bob about the main channel, and the current estimate $\hat{\mathbf{h}}$. Thus we define
\begin{equation}
    \Gamma = \sum_{n=1}^N \left(\left|\hat{h}_n^{(\textsc{AB})}\right|- \left|\hat{h}_n\right|\right).
    \label{eq:phi}
\end{equation}

Using such a simple modulus comparison alone results in a poor performance.
However, we can use both the criterion based on the LLR and that based on the modulus: only if both these conditions are met, then Bob accepts the message.
The verification condition can be therefore written as
\begin{equation}
 \begin{cases} \Psi \leq \theta, -\epsilon \leq \Gamma \leq \epsilon : & \mbox{decide for }\mathcal{H}_0, \\
 \mbox{else }: & \mbox{decide for }\mathcal{H}_1,
\end{cases}
\label{eq:sec_ver}
\end{equation}
where $\epsilon$ is a sufficiently small threshold. In the ideal case of absence of noise in phase 1, $\epsilon$ should be equal to zero. Since we are considering a realistic scenario affected by both noise and fading, we must allow $\epsilon$ to be greater than zero in order to allow Bob to accept messages coming from Alice.

The probability of false alarm can be therefore defined as the probability that at least one of the two conditions is not verified when the sender is Alice (hypothesis $\mathcal{H}_0$), i.e.
\begin{equation}
    P_{\textsc{FA}} = 1 - P\left[\Psi \leq \theta, -\epsilon \leq \Gamma \leq \epsilon | \mathcal{H}_0 \right],
    \label{eq:Pfa2}
\end{equation}
while the probability of missed detection can be written as
%The probabilities of false alarm and missed detection can then be written as
%\begin{equation}
%    P_{\textsc{FA}} = P\left[\Psi > \theta ,\Phi > \epsilon | \mathcal{H}_0 \right],
%\label{eq:Pfa2}
%\end{equation}
\begin{equation}
    P_{\textsc{MD}} = P\left[\Psi \leq \theta, -\epsilon \leq \Gamma \leq \epsilon | \mathcal{H}_1 \right].
\label{eq:Pmd2}
\end{equation}

%For their estimation we resort to Monte Carlo simulations.
%\mb{Qui non capisco cosa vogliamo dire: noi usiamo simulazioni Monte Carlo, however, per trovare una espressione analitica abbiamo bisogno di conoscere la distribuzione di $\Gamma$... ma se usiamo Monte Carlo l'espressione analitica a cosa ci serve? Poi pero diciamo che si puo calcolarla tramite la distribuzione di Hoyt... quindi il revisore si chiedera perche non lo facciamo... ?! Riscriverei in modo di generare meno dubbi.}
%However, we need to know the distributions of $\Gamma$ to find an analytic expression for the probabilities. 
%$\Gamma$ is computed as the difference between the modulus of two complex normal variables (whose real and imaginary parts are jointly normal).
Being $\Gamma$ computed as the difference of the modulus of two complex normal random variables, which follow Hoyt distribution \cite{Hoyt1947}, its \ac{c.d.f.} can be evaluated as in \cite[eq. (8)]{Paris2009}. However, a closed form expression for the joint probability distribution of $\Psi$ and $\Gamma$ is not known, and for the probabilities estimation we resort to Monte Carlo simulations.

\subsubsection{Thresholds optimization}
%In order to find the thresholds $\theta$ and $\epsilon$, we look for their joint optimal values that minimize the probability of MD, i.e.
%\begin{equation}
%   \bar{P}_{\textsc{MD}} =  \arg \min_{\theta, \epsilon} P\left[\Psi \leq \theta, -\epsilon \leq \Gamma \leq \epsilon | \mathcal{H}_1 \right],
%\end{equation}
%under the constraint of a fixed $P_{\textsc{FA}}$.

In order to find the optimal values of the thresholds $\theta^*$ and $\epsilon^*$, we look for their joint values which minimize the probability of MD, i.e.
\begin{equation}
  (\theta^*,\epsilon^*) = \arg\min_{\theta, \epsilon} P\left[\Psi \leq \theta, -\epsilon \leq \Gamma \leq \epsilon | \mathcal{H}_1 \right],
\end{equation}
under the constraint of a fixed $P_{\textsc{FA}}$.
We exploit an optimization procedure based on dichotomic search, that computes the couples $(\theta, \epsilon)$ which satisfy the constraint imposed on the $P_{\textsc{FA}}$ and then select the one that give the minimum $P_{\textsc{MD}}$.

Since we do not have a closed form expression for the probability of MD, we exploit an optimization procedure based on dichotomic search, that computes the couples $(\theta, \epsilon)$ which satisfy the constraint imposed on the $P_{\textsc{FA}}$ and then select the one that give the minimum $P_{\textsc{MD}}$.

\subsubsection{Attack strategy}
Regarding the LLR plus modulus comparison criterion, finding an optimal attack requires further considerations.
We start looking for a possible attack for the modulus comparison strategy, denoting it in the following as \textit{modulus attack} for brevity.
%For sake of simplicity, we suppose for the moment that Eve can observe transmissions from Alice to Bob, but not the channel between her and Bob. 
For sake of simplicity, we suppose that Eve can estimate the channel between Alice and herself, but not the channel between her and Bob.
%\mb{Qui mescoliamo l'osservazione delle trasmissioni con quella del canale. Poi diciamo Alice-Bob e Eve-Bob che non e simmetrico. Allora direi piuttosto che Eve puo stimare il canale tra Alice e se stessa, perche Alice trasmette, mentre non puo farlo per il canale tra lei e Bob, perche Bob non trasmette. Mi pare piu chiaro, no?}
Bob in fact receives messages from Alice but is not expected to transmit enough messages to allow the attacker to extract useful information on $\mathbf{h}^{(\textsc{EB})}$. Moreover, we assume that Eve is able to perfectly estimate $\mathbf{h}^{\textsc{(AE)}}$, i.e. $\sigma_{\textsc{AE}}^2 = 0$.

In order to make $\left|\mathbf{g}\right|$ more similar to $\left|\hat{\mathbf{h}}^{(\textsc{AB})} \right|$ Eve can forge 
\begin{equation}
    g_n = 
    \frac{\hat{\mathbf{h}}^{(\textsc{AE})}}{\rho_{\textsc{AE}}}, \mbox{for } \rho_{\textsc{AE}} \neq 0. 
\label{eq:att_mod}
\end{equation}

When we consider a decision strategy based on both \ac{LLR} and modulus comparison, however, attacks based on \eqref{eq:att1} and \eqref{eq:att_mod} are no longer optimal. 
In this case, the best attack strategy, which we will denote as \textit{combined attack} for brevity, in fact consists in writing the vector $\mathbf{g}$ as
\begin{equation}
g_n = \rho_{\textsc{\textsc{AE}}}^x \hat{h}_n^{(AE)},
    \label{eq:att3}
\end{equation}
where $x \in [-1,1]$. 
This requires finding a trade-off between the modulus attack and the \ac{LLR} attack. In fact, $x = -1$ corresponds to the best attack to the modulus comparison method, while $x = 1$ corresponds to the best attack strategy to the \ac{LLR}.
Finding the optimal value of $x$ corresponds to solve the problem
\begin{equation}
%\begin{split}
     x = \arg \max_{x \in [-1,1]} P\left[\Psi(x) \leq \theta, \Gamma(x)^2 \leq \epsilon^2 | \mathbf{g} = \rho_{\textsc{AE}}^x \mathbf{\hat{h}}^{(\textsc{AE})} \right],
    \label{eq:opt}
%\end{split}
\end{equation}
i.e. to find the value of $x$ that gives the highest probability of missed detection.
In Tab. \ref{tab:x_val} we report the optimal values of $x$ obtained by solving \eqref{eq:opt} through numerical methods, with $P_{\textsc{FA}} = 10^{\textsc{-4}}$ and different values of sub-carriers $N$ and time correlation $\mathbf{\alpha}$.
%We solve \eqref{eq:opt} through numerical methods, and in Tab. \ref{tab:x_val} we provide the resulting optimal values of $x$ fixing $P_{\textsc{FA}} = 10^{\textsc{-4}}$, for different numbers of sub-carriers $N$ and values of the time correlation parameter $\mathbf{\alpha}$. 
We observe that $x$ depends on the value of $\rho_{\textsc{AE}}$ and that, in the case of a single sub-carrier, the LLR attack represents the optimum solution for Eve even when Bob adopts the combined test.
%When $\rho_{\textsc{AE}} = 1$ Eve is in Bob's same position, thus she has his same estimate of \eqref{eq:2}, and an attack strategy is indeed unnecessary \textcolor{blue}{[...]}. 
%The corresponding value of $x$ is hence not reported in the Tab. \ref{tab:x_val}.
However, since the attacker does not know exactly the values of $\boldsymbol{\alpha}$, her most conservative choice is to suppose that all the entries of $\boldsymbol{\alpha}$ are equal to $1$. In fact, Eve is in the worst condition in absence of fading, since Bob lets fewer messages to be accepted as authentic.
%and the probability of MD decreases.  

\begin{table}[ht]
\begin{center}
\caption{Example of optimal values of the parameter $x$ for different values of $\alpha$, $\rho_{\textsc{AE}}$ and $N$. \label{tab:x_val}}
\begin{tabular}{|@{\hspace{0.5mm}}c@{\hspace{0.5mm}}|@{\hspace{0.5mm}}c@{\hspace{0.5mm}}|c|c|c|c|c|c|c|c|c|c|}
\hline
%\rule[-2mm]{0mm}{0.6cm}
\multirow{2}*{$\alpha$} & \multirow{2}*{$N$} & \multicolumn{10}{c|}{$\rho_{\textsc{AE}}$} \\
\cline{3-12}
 & & 0.1 & 0.2 & 0.3 & 0.4 & 0.5 & 0.6 & 0.7 & 0.8 & 0.9 & 1 \\
\hline
\multirow{3}*{1} & 1 & 1 & 1 & 1 & 1 & 1 & 1 & 1 & 1 & 1 & 1  \\
\cline{2-12} 
& $3$ & 0.9 & 1 & 1 & 1 & 1 & 1 & 1 & 1 & 1 & 1  \\
\cline{2-12}
& $6$ & 0.7 & 1 & 0.8 & 1 & 0.9 & 0.9 & 1 & 1 & 1 & 1  \\
\hline
\multirow{3}*{0.9} & 1 & 1 & 1 & 1 & 1 & 1 & 1 & 1 & 1 & 1 & 1  \\
\cline{2-12} 
& $3$ & 0.8 & 0.9 & 1 & 1 & 1 & 1 & 1 & 1 & 0.9 & 1 \\
\cline{2-12}
& $6$ & 0.6 & 0.8 & 0.9 & 0.9 & 0.9 & 0.9 & 0.9 & 0.9 & 0.7 & 1 \\
\hline
\end{tabular}
\end{center}
\end{table}
%\vspace{-1mm}
A comparative security assessment of the decision methods discussed above is reported in Section \ref{sec:results}.

\section{Machine learning-based decision techniques}
\label{sec:class}

In this section we consider the use of machine learning techniques in order to perform authentication according to the protocol discussed in Section \ref{sec:sys}. In particular, we consider \ac{OCNN} algorithms for the following reasons.
In the training phase, which corresponds to \textit{Phase I}, Bob receives a limited number of setup packets from Alice, while samples from Eve are not available, thus making the choice of a binary classification algorithm unsuitable for this setting. \ac{OCC} instead proves to be more effective, since the training set contains only samples from the target class (or positive class), while for the other classes (or negative classes) there are no instances (or they are very few or do not form a statistically-representative sample of the negative concept) \cite{khan2009}. The main purpose of \ac{OCC} is to define decision limits around the target class, in order to classify new instances (new messages in our case) as internal or external. In the case of a sample classified as external, this is not associated to a specific class, but is just identified as not belonging to the positive class (and therefore as a member of the negative class). The goal of \ac{OCC} is then more complex with respect to traditional classification, with particular regard to the definition of decision parameters and features to use in order to better distinguish positive and negative instances.

Since in authentication schemes it is reasonable to assume that \textit{Phase I} is short and Bob receives a limited number of setup packets, it is also fair to assume to work with a small training set. 
Among classification algorithms (such as Support Vector Machines, Bayesian Networks, Neural Networks, etc.), \ac{NN} methods are characterized by a complexity which increases with the training set dimension, being them instance-based: the training phase is based only on the memorization of the training set, thus making their use particularly suitable in the discussed scenario. Classification of new instances happens by considering the labels of the ``nearest" samples of the training set, on the basis of specific parameters \cite{Shalev2014}. They are also versatile to the implementation of both binary and one-class classification, and \ac{OCNN} techniques can be easily derived from traditional \ac{NN} algorithms.

In general, a \ac{OCNN} algorithm works as follows \cite{Khan2018}: it finds the \textit{j} nearest neighbors $\left\{y_1, \cdots, y_j \right\}$ of the test sample $x$ in the target class, and the \textit{k} nearest neighbors $\left\{z_{i1}, \cdots, z_{ik} \right\}$ of the first \textit{j} neighbors; it evaluates the average distance $\bar{D}_{xy}$ between $\left\{D_{xy_1}, \cdots, D_{xy_j} \right\}$ and the average distance $\bar{D}_{yz}$ between $\left\{D_{{y_1}{z_{11}}}, \cdots, D_{{y_1}{z_{1k}}}, \cdots, D_{{y_1}{z_{k1}}}, \cdots, D_{{y_k}{z_{kk}}} \right\}$; $x$ is then considered as a member of the target class if
    \begin{equation}
    \frac{\bar{D}_{xy}}{\bar{D}_{yz}} < \theta_d.
    \label{eq:JKNN}
    \end{equation}
\ac{OCNN} methods can be categorized into four main categories (11NN, 1KNN, J1NN, JKNN), depending on which of the parameters $j$ and $k$ is fixed to 1. They differ in the number of nearest neighbors used to compute the decision threshold. 

\textit{Phase II} of the protocol shown in Section \ref{sec:sys} corresponds to the actual classification. A new instance is recognized as belonging to the positive class if its distance from the target class is below a threshold $\theta_d$, where $d(\cdot)$ represents a generic distance between two instances. In formulas
\begin{equation}
    f(x) = I\left(d(x) < \theta_{d}\right).
\label{eq:oneclass}
\end{equation}
$I(\cdot)$ represents an indicator function and $f(x)$ is the binary function expressing acceptance of the object $x$ into the target class (i.e. the training set $X$). This is not different from verification conditions \eqref{eq:threshold} and \eqref{eq:sec_ver}, although here the presence of the nearest neighbors is crucial for establishing the source of the packet, while it is not considered in the analytical tests. 

We consider the Euclidean distance as metric, defined as 
\begin{equation}
    d\left(\mathbf{\hat{h}}^{\textsc{(AB)}}, \mathbf{\hat{h}}\right) = \sqrt{\sum_{i=1}^{2N} \left(a_i(\mathbf{\hat{h}}^{\textsc{(AB)}}) - a_i(\mathbf{\hat{h}}) \right)^2},
\label{eq:eucl}
\end{equation}
where $\{a_1(x), \cdots , a_{2N}(x)\}$ is the feature vector collecting the attributes of a generic instance $x$. In our case attributes are represented by the real and imaginary parts of the estimates $\mathbf{\hat{h}}^{\textsc{(AB)}}$ and $\mathbf{\hat{h}}$ measured on each sub-carrier.
Parameters $j$, $k$ and $\theta_d$ are optimized using a $g$-fold cross-validation \cite{Stone1974}.
Optimal attack strategy still consists of the \ac{ML} estimate of Alice-Bob channel.
Differently from the statistical methods examined, \ac{OCNN} techniques have the advantage of not requiring the knowledge of any \ac{CSI} by the authenticator.

\section{Numerical results}
\label{sec:results}

In this section, we assess and compare the performance of statistical and machine learning-based decision methods under different system conditions and assumptions. For the sake of simplicity, and without loss of generality, we consider examples in which time-varying fading affects all channels in the same way, i.e. $\alpha_n = \alpha$ for $n = 1, \cdots, N$. The average signal-to-noise ratio (SNR) on channel estimates during both phases has been calculated as a function of the noise variance as $\textsc{SNR}^{(\textsc{I})} = 1/\sigma_{\textsc{I}}^2$ and $\textsc{SNR}^{(\textsc{II})} = 1/\sigma_{\textsc{II}}^2$.

\begin{figure}[ht]
\centering
\includegraphics[width=0.41\textwidth]{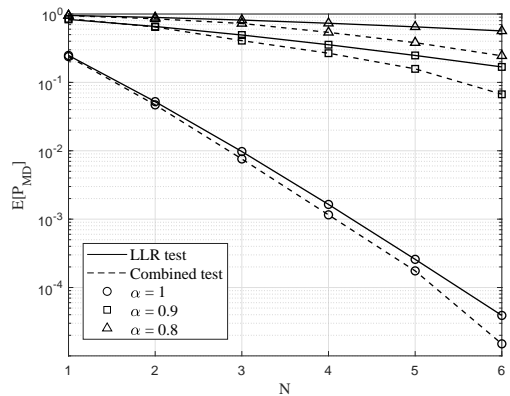}
\caption{Average MD probability $\mathbb{E}[P_{\textsc{MD}}]$ versus number of sub-carriers $N$, comparing three different test methods, for different values of $\alpha$, $\rho_{\textsc{AE}} = 0.1$, with $\textsc{SNR}^{(\textsc{I})} = 15$dB and $\textsc{SNR}^{(\textsc{II})} \rightarrow \infty$. \label{fig:confronto}}
\end{figure}

We first measure the security achieved by the statistical decision criteria presented in Section \ref{sec:dec_methods} in terms of average probability of MD. %For the sake of comparison, we consider another method based on the same euclidean distance of \eqref{eq:eucl} used by the classifiers, but in the analytical evaluation we do not take into account the nearest neighbor search included in the \ac{OCNN} algorithms. 
%\mb{La frase qui sopra non si capisce. Che consideriamo come paragone e perche? Riscrivere}
The probability of FA has been fixed equal to $10^{\textsc{-4}}$ for both the examined methods and the spatial correlation $\rho_{\textsc{AE}}$ has been set equal to 0.1, with the meaning of Eve very far from Bob's position. 
From the results in Fig. \ref{fig:confronto} it is evident how much the channel variability (represented by decreasing values of $\alpha$) degrades the performance of the system, with a significant increase in the probability of MD with respect to the flat fading case (i.e., $\alpha=1$).
Looking at the figure, we note that, with respect to the sole LLR test, the combined test helps Bob to enhance the performance of the scheme, and this becomes more and more evident for increasing numbers of sub-carriers. %The test based on euclidean distance achieves performance comparable to the LLR test with $\alpha = 1$, while it turns out to be the worst with smaller values of $\alpha$.

For \ac{OCNN} techniques, we consider a training set of $10^3$ samples and a data set of dimension $10^6$.
%\textcolor{red}{the same data set used for the Monte Carlo simulations, of dimension $10^6$}.
%By looking at Tab. \ref{tab:da_mettere}, 
In general none of the examined classifiers proves to achieve better performance than the others, although JKNN requires a higher computational cost. 
\begin{comment}
\begin{table}[ht]
\begin{center}
\caption{Probabilities of FA and MD achieved by different one-class classifiers choosing $\rho_{\textsc{AE}} = 0.1$ and $N = 3$. \label{tab:da_mettere}}
\begin{tabular}{|c|c|c|c|c|c|}
\hline
&  & 11NN & 1KNN & J1NN & JKNN  \\
\hline
\multirow{2}*{$\alpha = 1$} & $P_{\textsc{FA}}$ & $5\cdot10^{-5}$ & $<10^{-6}$ & $2.9\cdot10^{-5}$ & $<10^{-6}$ \\
\cline{2-6}
& $P_{\textsc{MD}}$ & $<10^{-6}$ & $<10^{-6}$ & $<10^{-6}$ & $<10^{-6}$  \\
\hline
\multirow{2}*{$\alpha = 0.9$} & $P_{\textsc{FA}}$ & 0.143 & 0.264 & 0.0935 & 0.461 \\
\cline{2-6}
& $P_{\textsc{MD}}$ & $<10^{-6}$ & $<10^{-6}$ & $<10^{-6}$ & $<10^{-6}$ \\
\hline
\multirow{2}*{$\alpha = 0.8$} & $P_{\textsc{FA}}$ & 0.303 & 0.668 & 0.757 & 0.825 \\
\cline{2-6}
& $P_{\textsc{MD}}$ & $<10^{-6}$ & $<10^{-6}$ & $<10^{-6}$ & $<10^{-6}$ \\
\hline
\end{tabular}
\end{center}
\end{table}
\end{comment}
In Tables \ref{tab:confronto_rho01} and \ref{tab:confronto_rho08} we show the performance achieved by \ac{OCNN} algorithms in comparison with the results obtained by applying the statistical methods to the same data set (we report the results obtained only by the classifier which results the best among the others in the considered cases) for two different values of the spatial correlation parameter $\rho_{\textsc{AE}}$. 
With $\rho_{\textsc{AE}} = 0.1$, we observe that \ac{OCNN} algorithms achieve excellent performance in a flat fading scenario, with probabilities of FA and MD less than $10^{-6}$, while when we consider more varying channels they tend to exhibit a higher probability of FA.\footnote{Probability $< 10^{-6}$ means that no error has been found over the entire data set.
In order to perform an analytical evaluation, these values have been considered equal to $10^{-6}$.
}  
This behavior is more evident when Eve is closer to Bob, i.e. when the value of $\rho_{\textsc{AE}}$ is higher. From Tab. \ref{tab:confronto_rho08} we observe that, while the probability of FA increases with the decreasing of $\alpha$ (and with the number of sub-carriers $N$), the probability of MD obtained by the classifier does not change with the time-varying fading.
This happens because, according to Phase I, the training set corresponds only to the initial state of the channel, thus remaining the same despite the time variability of the channel (in fact Bob is not aware of the value of $\alpha$ during the collection of the reference strings).
However, although statistical methods consider the role of the time-varying fading for the decision by means of the variance $\sigma^2$, which contains the time correlation parameter $\alpha$, \ac{OCNN} techniques result to be the best choice for low values of the spatial correlation.  %and in case of flat fading. 
%\textcolor{blue}{Ci sarebbe da aggiungere due parole perché il combined test va peggio dell'LLR quando $\rho$ e' alto e $\alpha<1$, ma non so di preciso come giustificarlo.}
We observe that the combined statistical test we have considered in our examples is efficient when Eve's and Bob's channels have low correlation, while the classic \ac{LLR} test alone may be the best solution in the case of highly correlated channels of Eve and Bob.

\begin{table}[ht]
\begin{center}
\caption{Average MD and FA probabilities obtained by different test methods, for different values of $\alpha$, with $\rho_{\textsc{AE}} = 0.1$, $\textsc{SNR}^{(\textsc{I})} = 15$dB and $\textsc{SNR}^{(\textsc{II})} = 20$dB. \label{tab:confronto_rho01}}
\begin{tabular}{|@{\hspace{0.3mm}}c@{\hspace{0.3mm}}||@{\hspace{0.3mm}}c@{\hspace{0.3mm}}|@{\hspace{0.3mm}}c@{\hspace{0.3mm}}|@{\hspace{0.3mm}}c@{\hspace{0.3mm}}|@{\hspace{0.3mm}}c@{\hspace{0.3mm}}|@{\hspace{0.3mm}}c@{\hspace{0.3mm}}|@{\hspace{0.3mm}}c@{\hspace{0.3mm}}|@{\hspace{0.3mm}}c@{\hspace{0.3mm}}|}
\hline
$\alpha$ & $N$ & 1 & 2 & 3 & 4 & 5 & 6 \\
\hline
\multirow{4}*{$1$} & $P_{\textsc{FA}}$ & 0.001 & $<10^{\textsc{-6}}$ & $<10^{\textsc{-6}}$ & $<10^{\textsc{-6}}$ & $<10^{\textsc{-6}}$ & $<10^{\textsc{-6}}$ \\
\cline{2-8}
& $P_{\textsc{MD}}$ (1KNN) & $\mathbf{<10^{\textbf{-6}}}$ & $\mathbf{<10^{\textbf{-6}}}$ & $\mathbf{<10^{\textbf{-6}}}$ & $\mathbf{<10^{\textbf{-6}}}$ & $\mathbf{<10^{\textbf{-6}}}$ & $\mathbf{<10^{\textbf{-6}}}$ \\
\cline{2-8}
& $P_{\textsc{MD}}$ (LLR) & 0.240 & 0.146 & 0.044 & 0.012 & 0.0029 & $6.9\cdot10^{\textsc{-4}}$ \\
\cline{2-8}
& $P_{\textsc{MD}}$ (comb) & 0.239 & 0.132 & 0.036 & 0.010 & 0.0019 & $5.5\cdot10^{\textsc{-4}}$  \\
\hline
\multirow{4}*{$0.9$} & $P_{\textsc{FA}}$ & 0.136 & 0.374 & 0.257 & 0.435 & 0.425 & 0.713 \\
\cline{2-8}
& $P_{\textsc{MD}}$ (1KNN) & $\mathbf{<10^{\textbf{-6}}}$ & $\mathbf{<10^{\textbf{-6}}}$ & $\mathbf{<10^{\textbf{-6}}}$ & $\mathbf{<10^{\textbf{-6}}}$ & $\mathbf{<10^{\textbf{-6}}}$ & $\mathbf{<10^{\textbf{-6}}}$ \\
\cline{2-8}
& $P_{\textsc{MD}}$ (LLR) & 0.320 & 0.064 & 0.040 & 0.0079 & 0.0034 & $3\cdot10^{\textsc{-4}}$ \\
\cline{2-8}
& $P_{\textsc{MD}}$ (comb) & 0.315 & 0.027 & 0.0096 & 0.0019 & 0.005 & $<10^{\textsc{-6}}$  \\
\hline
\multirow{4}*{$0.8$} & $P_{\textsc{FA}}$ & 0.358 & 0.668 & 0.602 & 0.812 & 0.848 & 0.964 \\
\cline{2-8}
& $P_{\textsc{MD}}$ (1KNN) & $\mathbf{<10^{\textbf{-6}}}$ & $\mathbf{<10^{\textbf{-6}}}$ & $\mathbf{<10^{\textbf{-6}}}$ & $\mathbf{<10^{\textbf{-6}}}$ & $\mathbf{<10^{\textbf{-6}}}$ & $\mathbf{<10^{\textbf{-6}}}$ \\
\cline{2-8}
& $P_{\textsc{MD}}$ (LLR) & 0.268 & 0.050 & 0.033 & 0.005 & 0.0018 & $1.1\cdot10^{\textsc{-4}}$ \\
\cline{2-8}
& $P_{\textsc{MD}}$ (comb) & 0.228 & 0.011 & 0.0027 & $3.9\cdot10^{\textsc{-5}}$ & $7\cdot10^{\textsc{-6}}$ &  $<10^{\textsc{-6}}$ \\
\hline
\end{tabular}
\end{center}
\vspace{-2mm}
\begin{center}
\caption{Average MD and FA probabilities obtained by different test methods, for different values of $\alpha$, with $\rho_{\textsc{AE}} = 0.8$, $\textsc{SNR}^{(\textsc{I})} = 15$dB and $\textsc{SNR}^{(\textsc{II})} = 20$dB. \label{tab:confronto_rho08}}
\begin{tabular}{|@{\hspace{0.5mm}}c@{\hspace{0.5mm}}||@{\hspace{0.5mm}}c@{\hspace{0.5mm}}|@{\hspace{0.3mm}}c@{\hspace{0.3mm}}|@{\hspace{0.3mm}}c@{\hspace{0.3mm}}|@{\hspace{0.3mm}}c@{\hspace{0.3mm}}|@{\hspace{0.3mm}}c@{\hspace{0.3mm}}|@{\hspace{0.3mm}}c@{\hspace{0.3mm}}|@{\hspace{0.3mm}}c@{\hspace{0.3mm}}|}
\hline
$\alpha$ & $N$ & 1 & 2 & 3 & 4 & 5 & 6 \\
\hline
\multirow{4}*{1} & $P_{\textsc{FA}}$ & 0.010 & $1.6\cdot10^{\textsc{-4}}$ & $<10^{\textsc{-6}}$ & $<10^{\textsc{-6}}$ & $<10^{\textsc{-6}}$ & $<10^{\textsc{-6}}$ \\
\cline{2-8}
& $P_{\textsc{MD}}$ (1KNN) & \textbf{0.206} & \textbf{0.110} & \textbf{0.041} & \textbf{0.018} & \textbf{0.012} & \textbf{0.005} \\
\cline{2-8}
& $P_{\textsc{MD}}$ (LLR) & 0.392 & 0.326 & 0.319 & 0.183 & 0.099 & 0.052 \\
\cline{2-8}
& $P_{\textsc{MD}}$ (comb) & 0.391 & 0.331 & 0.279 & 0.171 & 0.078 & 0.027  \\
\hline
\multirow{4}*{$0.9$} & $P_{\textsc{FA}}$ & 0.540 & 0.799 & 0.863 & 0.877 & 0.817 & 0.865 \\
\cline{2-8}
& $P_{\textsc{MD}}$ (1KNN) & \textbf{0.206} & 0.110 & 0.041 & 0.018 & 0.012 & 0.005 \\
\cline{2-8}
& $P_{\textsc{MD}}$ (LLR) & 0.221 & \textbf{0.045} & \textbf{0.016} & \textbf{0.008} & \textbf{0.008} & \textbf{0.003} \\
\cline{2-8}
& $P_{\textsc{MD}}$ (comb) & 0.353 & 0.127 & 0.060 & 0.032 & 0.035 & 0.021  \\
\hline
\multirow{4}*{$0.8$} & $P_{\textsc{FA}}$ & 0.739 & 0.928 & 0.970 & 0.983 & 0.982 & 0.991 \\
\cline{2-8}
& $P_{\textsc{MD}}$ (1KNN) & 0.206 & 0.110 & 0.041 & 0.018 & 0.012 & 0.005 \\
\cline{2-8}
& $P_{\textsc{MD}}$ (LLR) & \textbf{0.160} & \textbf{0.027} & \textbf{0.007} & \textbf{0.003} & \textbf{0.002} & $\mathbf{6.1\cdot10^{\textsc{-4}}}$\\
\cline{2-8}
& $P_{\textsc{MD}}$ (comb) & 0.258 & 0.067 & 0.029 & 0.021 & 0.017 & $9.7\cdot10^{\textsc{-4}}$ \\
\hline
\end{tabular}
\end{center}
\end{table}

\section{Conclusion}
\label{sec:conclusion}
We have considered the problem of physical layer authentication based on channel estimates in the case of time-varying fading channels. In order to overcome the loss in terms of security coming from this phenomenon, we have proposed different decision methods for an authenticator to establish who is the source of the message and detect forged channel estimates. These methods are based on statistical criteria and \ac{OCC} techniques. We have shown that, although the training phase ignores the presence of the fading, \ac{OCNN} algorithms prove to offer better security performance in most of the examined cases, while statistical methods maintain some advantage when a strong spatial correlation between the attacker and the authenticator exists. 
%As a future work, we will study how to adapt the training phase of one-class classification taking into account the presence of fading.

\bibliographystyle{IEEEtran}
\bibliography{Archive.bib}

\end{document}